\documentstyle[preprint,aps,graphicx]{revtex}
\begin{document}
\draft
\title{Conductivity Oscillations in Current-Induced Metastable States in
Low-Doped Manganite Single Crystals}
\author{V. Dikovsky, Y. Yuzhelevski, V. Markovich, G. Gorodetsky,
and G. Jung\footnote{also with Institute of Physics, PAN, 02-668 Warszawa,
Poland}}
\address{Department of Physics, Ben Gurion University of the Negev,\\
P.O. BOX 653, 84105 Beer-Sheva, Israel}
\author{D. A. Shulyatev and Ya. M. Mukovskii}
\address{Moscow State Steel and Alloys Institute, 117936, Moscow, Russia}
\date{\today}
\maketitle
\begin{abstract}
Deterministic oscillations of current-induced metastable
resistivity in changing voltage have been detected in
La$_{0.82}$Ca$_{0.18}$MnO$_3$ single crystals. At low
temperatures, below the Curie point, application of specific bias
procedures switches the crystal into metastable resistivity state
characterized by appearance of pronounced reproducible and random
structures in the voltage dependence of the differential
conductivity. In certain bias range equally spaced broad
conductivity peaks have been observed. The oscillating
conductivity has been tentatively ascribed to resonances in a
quantum well within the double tunnel barrier of intrinsic
weak-links associated with twin-like defect boundaries.
\end{abstract}
%\begin{multicols}{2}

Transport properties of perovskite manganites are strongly
influenced by external factors such as magnetic field, electric
fields, hydrostatic pressure, electromagnetic irradiation, and
transport current.\cite{dagotto} Spectacular manifestations of
electric current effects in perovskite manganites are abrupt,
several orders of magnitude strong, current enforced resistivity
jumps.\cite{jumps} The phenomena of resistivity switching in
low-doped La$_{1-x}$Ca$_{x}$MnO$_3$ (LCMO) crystals at low
temperatures can be seen as hall-mark of transitions into
current-induced metastable resistivity states.\cite{ourPRB} The
properties of metastable states depend primarily on the bias
history of the sample. Metastable states are characterized by
appearance of reproducible and random structures in the bias
dependence of the differential conductivity. The conductivity has
a pronounced random noise component, frequently of a non-Gaussian
character.\cite{FNL} In this paper we report on deterministic
periodic conductivity oscillations appearing in current-induced
metastable state of La$_{0.82}$Ca$_{0.18}$MnO$_3$ (LCMO) single
crystal in zero applied magnetic field.

Single crystals of La$_{0.82}$Ca$_{0.18}$MnO$_{3}$ were grown by floating
zone technique, as described in details elsewhere.\cite{crystal} For
transport measurements the as-grown sample has been cut into
$8\times3\times1.6$ mm$^3$ rectangular bars with the longest dimension
parallel to $<110>$ crystalline direction. The differential resistance of
the sample $R_d=dV/dI$ has been measured using a lock-in technique in a
four-point contact arrangement.

Current bias procedures that were employed to create metastable
resistivity states in our sample are illustrated in Fig.~1. The
LCMO crystal is initially cooled to low temperatures well below
the Curie temperature $T_c$ in zero applied magnetic field.
Application of the bias current at this stage traces smooth,
bell-shaped $R_d(I)$ characteristics shown with a dashed line. The
characteristics is fully reversible, provided the bias current
does not exceed some thresholds labelled (2) and (3). When the
applied current exceeds a threshold value the resistance abruptly
drops and the metastable resistivity state (MRS) is established.
Upon subsequent current decrease to zero followed by an increase
towards the negative threshold the distorted bell-like
characteristics of the MRS is traced (dash-dotted line). The MRS
$R_d(I)$ characteristics are fully reproducible provided the
current does not exceed any of the threshold points. The shape of
the MRS characteristics is very close to the one of the pristine
state (dashed line), nevertheless, the MRS bell is slightly
distorted and contains clearly visible structures, in particular
at currents close to the thresholds. When the bias current applied
in a direction opposite to that which triggered the preceding
transition to the MRS exceeds the threshold value, in the case
illustrated in Fig.~1 the threshold (3), sample resistance jumps
more pronouncedly and the new metastable low resistivity state
(LRS) is established. The resistance of the LRS, solid line in
Fig.~1, is markedly different from previously observed pristine
and MRS characteristics. Instead of the bell-shaped form we have a
complex curve containing a rich peak structure, pronounced broad
minimum at low bias, and a zero-bias anomaly in the form of a
small local resistivity maximum at zero bias.  The preceding MRS
state can be restored by driving the bias current in the positive
direction above the threshold (2). Otherwise, for all currents
contained between (1) and (2) the sample remains in the LRS. For
current ramp exceeding both thresholds a strong hysteretic
behavior is observed. The resistivity seen at the
positive-to-negative current direction follows the MRS
characteristics, while the LRS characteristics is being traced
upon current return from negative to positive direction.

Obviously, figure 1 illustrates just one example of many possible
realizations of the metastable resistivity in our sample. Upon
increasing the range of the current excursions above the threshold
points (2) and (3), the LRS gradually evolves. The zero bias
resistance decreases and conductivity maxima move towards higher
voltages. Eventually, upon reaching certain high bias threshold
the resistance abruptly jumps again and one more metastable state
referred to as very low resistivity state (VLRS) is established.
The general features of the LRS characteristics are preserved in
the VLRS, in particular the resistivity minimum at zero bias and
pronounced conductivity peaks. The VLRS is relatively stable and
possesses a long-term memory of its resistivity. The memory
survives even the thermal cycling to room temperature and
maintaining the sample for few days at 300 K.\cite{ourPRB}

Fig.~2 shows the temperature evolution of the pristine and VLRS
resistance during a slow warm-up from 77 K to room temperatures.
The solid line was measured with a sample maintained previously
only in the pristine state. The arrow indicates the Curie
temperature established from the independent magnetization
measurements. The Curie point coincides with the local resistance
maximum and no significant hysteresis is observed between the
cooling and heating run. The dash-dot line represents the
resistance of a sample in which the VLRS has been enforced by
current procedures at 77 K. The VLRS constitutes the lower limit
of the observable evolution of the metastable resistivity and thus
the area contained between solid and dot-dash lines in Fig.~2
represents the range of available metastable resistivity. The
evolution of  $R_d(V)$ characteristics with increasing range of
the current  sweep is illustrated in the inset to Fig.~2. Note
that the resistance of all metastable states starts to converge
slightly above $T_c$ at a temperature which can be associated with
the Jahn-Teller transition.\cite{ourPRB,diagram}

Metastable resistivity states are characterized by stochastically
fluctuating conductivity.\cite{FNL} However, we have found that
the peak structure associated with the broad zero-bias
conductivity maximum remains very reproducible, in particular when
plotted as a function of the bias voltage rather than the current.
We have previously suggested that the $R_d(V)$ structures can be
associated with the tunnel character of the low temperature
conductivity.\cite{ourPRB} Tunnel mechanism starts to dominate
when metallic ferromagnetic percolation path becomes more and more
interrupted by intrinsic tunnel weak-links. The peak structure
likely reflects the shape of the density of states in the
ferromagnetic areas coupled by the tunnel barrier, or in the
barrier itself, as it will be discussed elsewhere. In this paper
we concentrate on the phenomena of deterministic conductivity
oscillations found in voltage dependence of the dynamic
resistivity  in a metastable states.

Fig.~3 illustrates the voltage dependence of the conductivity of
current enforced MRS. At the bias close to the thresholds  the
conductivity has a "noisy" character. However, the enlarged view
of the encircled part of the characteristics, see inset, shows
that the "noise" constitutes in fact a series of equally spaced
peaks. The peak structure is highly reproducible as demonstrated
in Fig.~3 by coincidence of the peak positions in two records
measured separately under an increasing and decreasing current.
The peak voltages determined for the negative and positive branch
of a $R_d(V)$ characteristics are plotted in Fig.~4 as a function
of the peak number. The data in both directions are well fitted by
the linear dependence with the same slope. Good linear fit proves
that the observed oscillations are indeed periodic. From the slope
of the linear fit we determine that the voltage periodicity of the
conductivity oscillations in the MRS is $\Delta V=39 \pm 1 $mV.
The deterministic and periodic nature of the peak structure has
been confirmed by the Fourier analysis. The Fourier spectrum shown
in the inset was obtained by averaging several $R_d(V)$ recordings
obtained during subsequent current sweeps. The spectrum exhibits a
broad peak centered at the voltage corresponding to the period
determined by fitting procedures.

At low temperatures, below the the metal-insulator transition the
phase separated state becomes the stable ground state of the
low-doped LCMO compound and the ferromagnetic phase forms
percolating conducting paths. With decreasing temperature the
volume of the ferromagnetic phase increases. However, as indicated
by the resistivity upturn in Fig.~2 and by the nonlinear character
of the I-V characteristics, the low temperature conductivity  is
dominated by tunnel mechanism. Although the absolute proof of the
tunnel conductivity cannot be provided, we remaind the reader that
we have previously shown that voltage and temperature dependencies
of the resistance can be very well fitted to the Glasman and
Matveev (GM) model of indirect tunnelling.\cite{ourPRB,GM} The
nature of the involved intrinsic tunnel junctions remains still an
open question. Growth of the insulating phase that interrupts the
percolation path at low temperatures is one of the possible
scenarios. Nevertheless, in the existing literature there is no
consensus about the existence of the low temperature phase
transition to the insulating state at the doping range of our
samples.\cite{dagotto} On the other hand, using the magneto-optic
(MO) techniques we have always detected in our crystal  at
temperatures below $T_c$ a regular pattern of stripe-like domains
with alternating level of magnetization. The MO contrast of the
stripes depends on the applied field, the high and low
magnetization inverts coherently upon inverting the field
direction, but neither the strip positions nor their size, in the
range of $50-200 \mu$m, do not change with changing field and
temperature. On subsequent cooling cycles the details of the
pattern change but the overall features remain invariant. This
indicates the structural origin of the magnetic patterns.

The most probable source of stripe-shaped domains can be
occurrence of twins, rotation defects or regular grain boundaries
in the investigated crystal. Manganite CMR crystals are known to
contain twin-like structural defects which are strongly correlated
with magnetic domains.\cite{twins,twins07,vv} Moreover, in
low-doped LCMO such defects are capable of  pinning and preventing
the motion of the magnetic domain walls.\cite{magpin} We note at
this point that the observed MO domain size and direction within
the crystal is not constant.  The domain structure does not
reflect therefore, the presence of coherent twins similar to those
in manganites with charge-ordered low temperature phase
transition.\cite{podzorov} This is due to the absence of intrinsic
martensitic transition in 0.18 LCMO system. Intrinsic tunnel
weak-links in low carrier density manganites  are prone to be
localized at structural defects. Band bending effects in the
ferromagnetic metallic phase in the vicinity of a twin or grain
boundary  result in creation of depletion layers acting as
localized insulating tunnel barriers.\cite{grossrev,gross} The
band bending, and consequently the intrinsic junction conductance
depends on the difference in magnetization between adjacent
domains. Due to the stress associated with the lattice mismatch, a
few nm thick strongly disordered accommodation layer containing
arrays of dislocations is formed within the
boundary.\cite{grossrev} The stress field within the accommodation
layer result in local depression of the Curie temperature and
creation of a interfacial conducting layer contained between two
insulating tunnel barriers on both sides of the twin
boundary.\cite{grossrev,gross} Consequently, the intrinsic
junction has a complex structure of a double barrier tunnel
junction.

It has been shown theoretically and confirmed experimentally that a
nonmagnetic interface layer, even within a single tunnel barrier magnetic
junction, behaves as quantum well leading to resonant enhancements of
conductivity whenever a resonant condition for a quantum well state is
fulfilled.\cite{zhang98:qwell:zhangJAP} In a double barrier magnetic tunnel
junction, quantum well layer resonances enable coherent tunnelling of
electrons through the entire junction resulting in strong enhancement of
the conductivity at resonant conditions.\cite{zheng99,wilcz} In the
approximation of the spin-polarized free electron model the quantum well
resonances occur whenever $k_Fa=n\pi$, where $a$ is the width of the
interface layer and $k_F$ the Fermi wave vector.\cite{zheng99,zhang97:vied}
This model predicts that the periodicity of conductivity oscillations  is
\begin{equation}
\Delta{V}\approx \frac{h}{4a^2m}\frac{h}{2e} . \label{period}
\end{equation}
By confronting the experimentally determined periodicity of conductivity
oscillations with the prediction of the model we obtain the width of the
interface disordered layer $a=3.1$ nm.

Let us point out that a regular network of grain boundaries can by itself
lead to resonances in a multi-barrier spin dependent system of
ferromagnetic quantum wells.\cite{zhang97:vied} This scenario is however
very unlikely due to macroscopic distances between the twin boundaries seen
in the magneto-optics images. Scattering processes during electron transit
between individual boundaries will prevent the formation of quantum well
states. Another alternative mechanism for conductivity oscillations assumes
existence of a small mesoscopic metallic grain within the tunnel barrier.
If the grain is small enough, the charging effect may lead to Coulomb
blockade of the tunnelling current.\cite{coulomb} However, the estimations
of the grain size and resulting periodicity at 77 K give unrealistic
values, far from those observed in the experiments, while the estimations
based on the double barrier grain boundary junction give very plausible
values for the disordered interface layer thickness. Therefore, despite the
fact that we cannot provide a direct proof for the validity of the proposed
mechanism, we believe that coherent spin polarized resonant tunnelling lies
behind the experimentally observed conductivity oscillations.

\acknowledgments  This research was supported by the grant 209/01 from
Israeli Science Foundation.

\begin{figure}
%\centerline{\includegraphics*[width=\hsize]{fig1.eps}}
\caption{Current procedures and resulting metastable resistivity
states. Current is first increased from (1) (dashed line =
pristine state) slightly above the threshold (2), decreased
towards zero, and further increased in the negative direction
(dash-dotted line = MRS) slightly above the threshold (3). The
solid line: subsequent current sweep (LRS) from (3) towards (2).}
\end{figure}

\begin{figure}
%\centerline{\includegraphics*[width=\hsize]{fig2.eps}}
\caption{Temperature dependence of the zero-bias dynamic
resistance during a slow warm-up procedure. Solid line: the
pristine state. Dash-dot line: the VLRS state. Area between the
lines contains all possible metastable states. Inset:
Characteristics 1$\div 3$ illustrate evolution of the metastable
low resistivity state with increasing range of the bipolar current
sweep.}
\end{figure}

\begin{figure}
%\centerline{\includegraphics*[width=\hsize]{fig3.eps}}
\caption{Conductivity vs. voltage in the MRS with pronounced
oscillations at bias close to the threshold. The encircled part of
the characteristics is shown in the inset for increasing and
decreasing current ramp.}
\end{figure}

\begin{figure}
%\centerline{\includegraphics*[width=\hsize]{fig4.eps}}
\caption{Voltage positions of the conductivity peaks as a function
of the peak number. Full symbols - positive current, open symbols
- negative current. Inset: Averaged Fourier spectrum of several
conductivity vs voltage records. Broad peak centered at the
voltage periodicity as determined from the peak analysis is
indicated by an arrow.}
\end{figure}

%\end{multicols}
%\end{document}.
\vfill\eject\rotatebox{90}{\includegraphics*[width=22truecm]{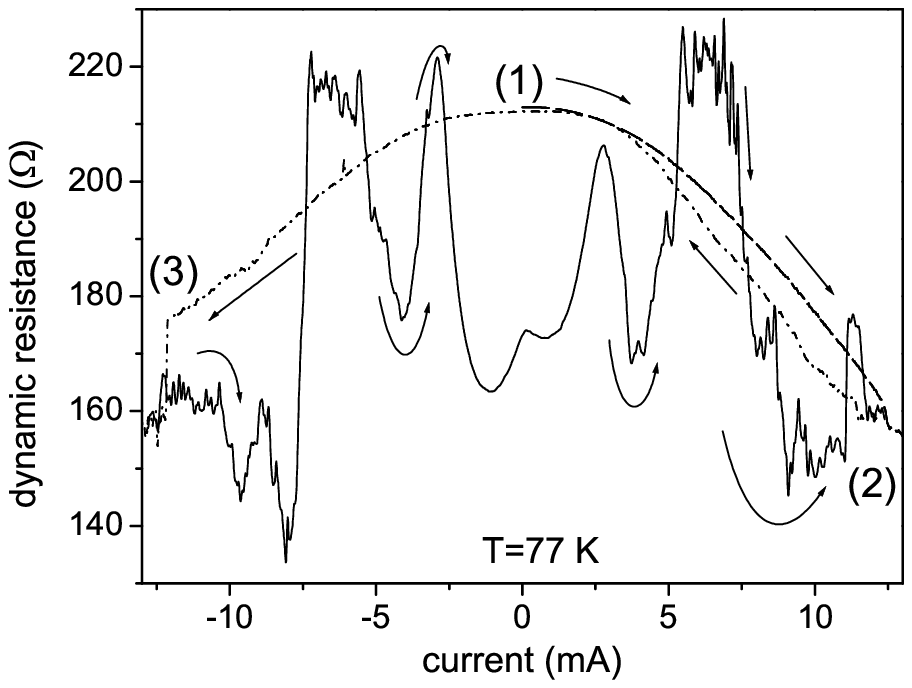}}
  {\it Fig. 1, Dikovsky et al., Conductivity Oscillations...}
 \vfill\eject\rotatebox{90}{\includegraphics*[width=22truecm]{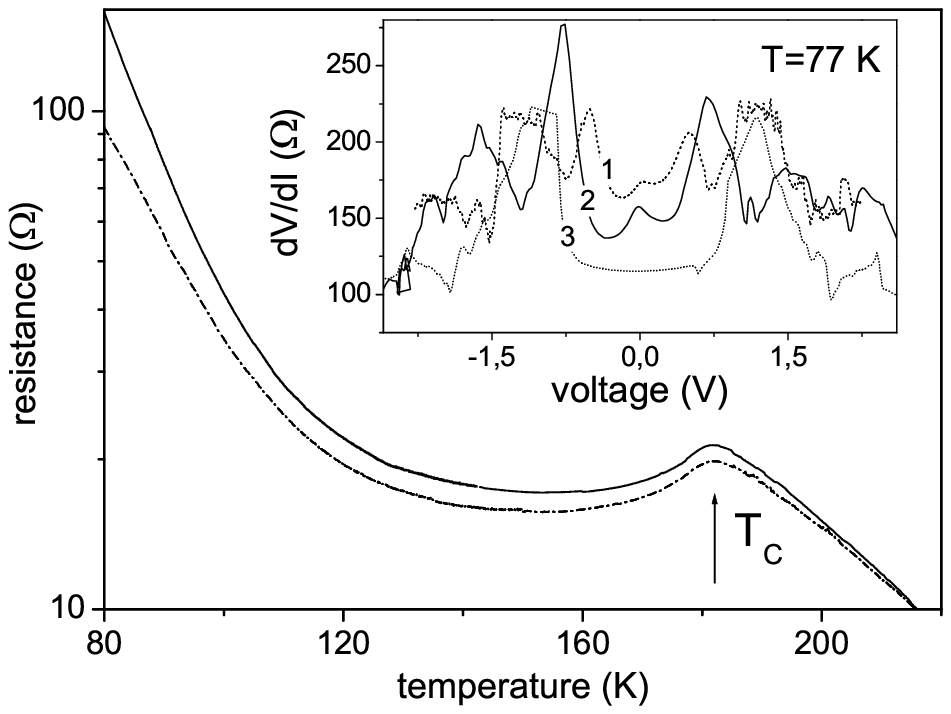}}
 {\it Fig. 2, Dikovsky et al., Conductivity Oscillations...}
 \vfill\eject\rotatebox{90}{\includegraphics*[width=22truecm]{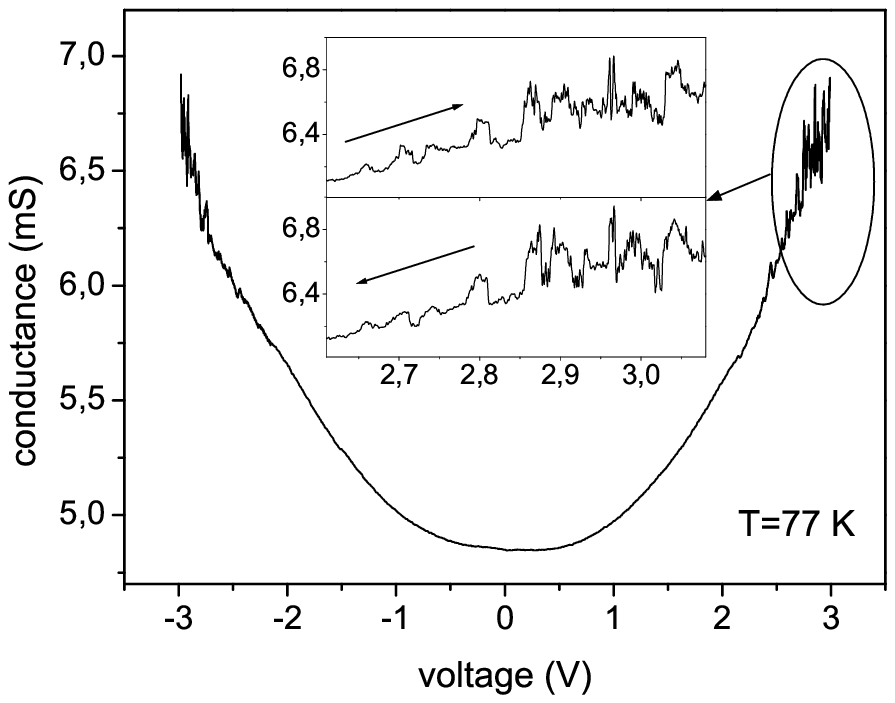}}
 {\it Fig. 3, Dikovsky et al., Conductivity Oscillations...}
 \vfill\eject\rotatebox{90}{\includegraphics*[width=22truecm]{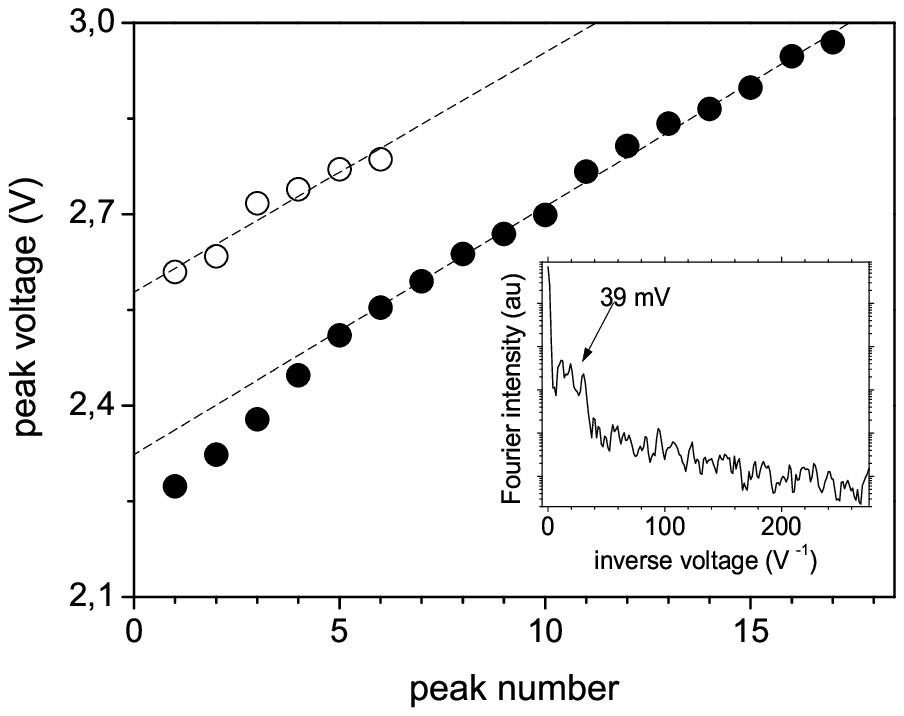}}
 {\it Fig. 4, Dikovsky et al., Conductivity Oscillations...}
\end{document}